\documentclass[twocolumn]{aastex631}

\usepackage{natbib}
\usepackage{subfigure}

\begin{document}

\title{The Impact of Local Stellar Radiation on Dwarf Galaxy Formation Around Milky Way Analogues Across Cosmic Reionization}

\author{Bocheng Zhu}
\affiliation{Key Laboratory for Computational Astrophysics, National Astronomical Observatories, \\Chinese Academy of Sciences, Beijing, China}
\affiliation{School of Physics and Laboratory of Zhongyuan Light, Zhengzhou University}

\author{Liang Gao}
\email{lgao@bao.ac.cn}
\affiliation{School of Physics and Laboratory of Zhongyuan Light, Zhengzhou University}
\affiliation{Institute for Frontiers in Astronomy and Astrophysics, Beijing Normal University, Beijing 102206, China}
\affiliation{Key Laboratory for Computational Astrophysics, National Astronomical Observatories, \\Chinese Academy of Sciences, Beijing, China}



\begin{abstract}

We explore the effect of local stellar radiation on the formation and evolution of {  dwarf galaxies} around Milky Way (MW) analogues. Using five simulations from the Auriga project, both with and without local stellar radiation, we find that local stellar radiation, as a pre-reionization source, is {  highly} effective at photoionizing and heating the gas around the proto-MW analogues. As a result, the formation of surrounding dwarf galaxies in {  dark matter halos with masses} below approximately $10^{9.5}\,\mathrm{M_{\odot}}$ are significantly suppressed. {  After reionization}, the intensity of local stellar radiation eventually becomes comparable to the ultraviolet background (UVB){ . Consequently,} the impact of local stellar radiation on the surrounding dwarf galaxy formation decreases with decreasing redshift and {  nearly} vanishes after redshift $z=4$. At present day, the bright satellite population in the simulations with and without local stellar radiation is nearly identical. While our simulations do not have {  sufficient} resolution to resolve the faintest satellite galaxies which are most prone to the local stellar radiation, we use {  the} accreted galaxy mass function to assess the impact and find that the reduction in the faintest satellite is around $13$ percent in the presence of local stellar radiation, {  but this difference is within $\sim2\sigma$ of the Poisson uncertainty and thus not statistically significant.}

\end{abstract}


\defcitealias{zhu2024}{Paper I}
\section{Introduction} \label{sec:intro}

Understanding the properties of dwarf galaxies provides a crucial testbed for different dark matter (DM) models, such as warm DM or fuzzy DM, which exhibit notable differences at {  small scales} \citep{klypin1999, bode2001, lovell2012, viel2013, weinberg2015, bullock2017, delpopolo2017, tulin2018, Mocz2019, may2021}. Although the underlying cosmological model influences {  small-scale} structure, astrophysical processes like reionization also play significant roles \citep[][etc.]{benson2002, okamoto2008, Okamoto2010, sawala2016, dashyan2018, Koudmani2021, liu2024}. For example, the well-known missing satellites problem can {  possibly be} resolved by introducing the astrophysical processes \citep{benson2002}. The missing satellites problem arises from the {  mismatch between the abundance predicted by N-body simulations} based on the $\Lambda$CDM model and the observed satellite galaxies of the Milky Way (MW) \citep{klypin1999, bullock2017}. This problem can be alleviated by considering cosmic reionization during which ultraviolet (UV) photons emitted by stars photoionize the neutral gas, {  heat gas} and inhibit further star formation in low mass dark matter halos \citep[ultraviolet background, hereafter UVB,][]{haardt96, faucher09, haardt12, faucher20}. With UVB, theoretical works using semi-analytical models (SAMs) {  have successfully reproduced} the abundance and properties of the dwarfs \citep{benson2002}.

Extensive studies have been carried out to explore {  the} impact of UVB on galaxy formation in {  low-mass} DM halos. For example, \citet{okamoto2008, Okamoto2010} show that UVB can heat the gas, {  causing it to escape} from the DM halo with low circular velocity and prevent them from forming stars. \citet{sawala2016} further confirms that reionization can reduce the number of dwarfs below a certain DM halo mass. \citet{Benitez2017} found that the UVB can photoionize the HI gas at the {  outskirts} of {  dwarf galaxies} while the HI gas at the center of dwarfs will not be affected. 

Most {  of the} studies discussed above assumed that {  the} reionization process starts and completes instantaneously at redshift $6$ and treated the UVB as a spatially uniform, time-dependent radiation field. However, neither the cosmic reionization is instantaneous nor the ``real'' UVB  is uniform. The cosmic reionization is a long process{  ;} once the very first stars and galaxies form, they {  begin} to produce {  a} radiation field {  that} influence galaxy formation in their surrounding DM halos. Furthermore{ ,} the radiation field is stronger near the massive galaxies. This excess radiation may also affect the formation and evolution of {  nearby dwarf galaxies}. 

Several studies have examined the impact of local radiation feedback on central galaxies \citep[][etc.]{hopkins12, Roskar14, kannan14, ceverino2014, obreja19, hopkins20, zhu2024}. However,  studies of the effect of local radiation on nearby dwarf galaxies around Milky Way-sized (MW-sized) galaxies are still scarce. \citet{ocvirk2014} applied radiative transfer (RT) code to post-process simulation data from the Constrained Local UniversE Simulations \citep[CLUES; ][]{clues2010} and found that reionization for the satellites occurs inside-out. \citet{Ocvirk2016} implemented a full RT in their Cosmic Dawn (CoDa) project, which is a cosmological zoom-in simulation project of the Local Group. They demonstrated that local stellar radiation substantially suppresses star formation in halos with masses below approximately $2 \times 10^9~\mathrm{M_{\odot}}$. Further analysis by \citet{Dawoodbhoy2018} on the CoDa project data indicated that the star formation in halos lower than $10^9~\mathrm{M_{\odot}}$ is suppressed by the local stellar radiation.

In this paper, we analyze simulation data from the Auriga project using an effective RT method \citep[][hereafter \citetalias{zhu2024}]{zhu2024} to explore the impact of local stellar radiation on dwarf galaxies near proto MW-sized galaxies.

The structure of this paper is as follows. In Section \ref{sec:methods}, we briefly introduce the simulation suites, the galaxy formation model and the model of the local stellar radiation implemented in the simulations. In Section \ref{sec:results}, we present and discuss our main results. Finally, we summarize our findings in Section \ref{sec:summary}.

\section{Methodology} \label{sec:methods}

\subsection{The Auriga-ERT simulations}

Following \citetalias{zhu2024}, the simulations used in this work are based on the Auriga project, which is a suite of high-resolution magnetohydrodynamical cosmological zoom-in simulations of MW {  analogues} \citep{grand17},  which was recently made publicly available \citep{Grand2024}. The simulation is performed with \textsc{arepo} code \citep{springel10}, which is {  an} N-body and moving-mesh magnetohydrodynamical code. Due to its quasi-Lagrangian nature, the \textsc{arepo} code provides an adaptive resolution and {  maintains} Galilean invariance for fluid dynamics.

{  Unlike} the original Auriga project, we incorporate the galaxy formation model from the IllustrisTNG project \citep{weinberger17, pillepich18}, with star formation and stellar wind model based on \citep{springel03}. The gas cells in the simulation convert to star-forming gas when their density exceeds a threshold and temperature is lower than a temperature criterion calculated with a two-phase ISM model. The star formation rate (SFR) is determined by this two-phase ISM model. The power of the stellar wind is then obtained based on SFR and is injected in a non-local form. The black holes (BHs) feeding and feedback model is described in \citet{weinberger17}. BHs are treated as sink particles and the BH accretion rate is calculated with the Bondi-Hoyle accretion formula. The BH feedback is separated into two modes based on the BH accretion rate. At a high accretion rate, the BH feedback is treated as a thermal mode, {  and} the feedback energy is injected isotropically in the form of thermal energy. While at a low accretion rate, the BH feedback is treated as a kinetic mode, and the feedback energy is injected in the form of kinetic energy. The direction of the kinetic outflow is randomly assigned in each feedback event, with kinetic energy injected along this direction.

The Auriga project selected 30 MW-size halos from the EAGLE dark matter only (DMO) simulation { ,} which has a box size 100 c${\rm Mpc}/h$ on a side,  and re-simulated them with ``zoom-in'' technique \citep{schaye2015, crain15}. To investigate the impact of local stellar radiation on the dwarfs near {  central} galaxies, we {  selected} $5$ of $30$ halos in the Auriga simulations and re-simulated them with and without the local stellar radiation described below. The selection criterion is based on their disk morphologies and merger states, ensuring a diverse range of formation histories, including extended stellar disks, compact disks, and a major merger case at $z=0$. This selection was made to capture variations in formation histories while maintaining numerical feasibility. The simulated galaxies with local stellar radiation {  are} referred to as StarRad simulations, while those without the local stellar radiation {  are} referred to as NoRad simulations.

\subsection{The model of the local stellar radiation}

The model of local stellar radiation used in the simulation suits is based on \citet{kannan14} with some modifications. The radiation field in the simulations originates from the UV background (UVB) {  as well as star-forming} regions, old stars and BHs of galaxies. For the UVB, we use the spectral energy distribution (SED) given by \citet{faucher20}, which updates previous models to consider non-uniform shielding effects. 

Radiation from {  star-forming} regions is modelled using the Binary Population and Spectral Synthesis (BPASS) v2.3 model \citep{eldridge17} for young stars, augmented with a power-law type emission for high mass X-ray binaries (HMXBs) and {  a multi-temperature Raymond-Smith model \citep{raymond1977}} for supernova remnants (SNR). For simplicity, instead of using the time-dependent SED, we use the SED of the stellar population in {  the} BPASS v2.3 model at an age equal to 10 Myr as the radiation emitted from the young stars. We also assume that the radiation from {  star-forming} region will be absorbed when they propagate out. {  We calculate the escape fraction of the radiation following \citet{kannan14}, which is based on \citet{cantalupo10}, and is described as a function of frequency $f^{\nu}_{\rm esc}$:}
\begin{equation}
f^\nu_{\rm esc} = f^{\rm LL}_{\rm esc}+(1-f^{\rm LL}_{\rm esc})\,{\rm e}^{-\tau_{\nu}}.
\end{equation}
, where $\tau_{\nu}$ is the optical depth of neutral hydrogen in {  the} star-forming region. The escape fraction at the Lyman limit frequency $f^{\rm LL}_{\rm esc}$ is set to 5\%, which is based on the observation of local starburst \citep{Bergvall2006} and high-redshift LBG \citep{steidel2001, shapley2006}. Based on the model described above, the scaling relation between the luminosity of the young stars and the SFR is 
\begin{equation}
L_{\rm YS}\approx 1.4\times 10^{42} \frac{\rm SFR}{\rm M_{\odot}~yr^{-1}}~{\rm erg~s^{-1}}.
\end{equation}. 
The HMXBs are modeled using a power law with slope $\Gamma=2$ at $0.5-8.0\,\rm keV$ \citep{anderson13}. The scaling relation between SFR and the X-ray luminosity from the HMXBs is 
\begin{equation}
L_{\rm HMXB}\approx 1.4\times 10^{39} \frac{\rm SFR}{\rm M_{\odot}~yr^{-1}}~{\rm erg~s^{-1}}.
\end{equation}. 
For the radiation from SNR, we follow the model {  described in} \citet{cervino02}. Based on their work, 20\% of the SN energy will convert to the radiation of SNR. Thus, the corresponding scaling relation between the radiation of the SNR and SFR is 
\begin{equation}
L_{\rm SNR}\approx 9.5\times 10^{39} \frac{\rm SFR}{\rm M_{\odot}~yr^{-1}}~{\rm erg~s^{-1}}.
\end{equation}.  The SED of the SNR is 
\begin{equation}
    f_{\nu} = 0.65 f_{\rm RS}(0.76 {\rm keV}) + 0.175 f_{\rm RS}(0.23 {\rm keV}) + 0.175 f_{\rm RS}(1.29 {\rm keV})
\end{equation}
where $f_{\rm RS}(kT)$ is the Raymond-Smith model \citep{raymond1977}. 


The radiation from {  the} old stellar region includes the old stars and low-mass X-ray binaries (LMXBs). Similar to young stars, old stars also use {  the} BPASS v2.3 model with radiation properties based on a {  2-Gyr-old} stellar population. The scaling relation between the luminosity of old stars and the stellar mass is 
\begin{equation}
L_{\rm OS} \approx 5.5\times10^{40}\frac{M_{\star}}{10^{11}{\rm M_{\odot}}}~{\rm erg~s^{-1}}.
\end{equation}
The model of (LMXBs) is also based on \citet{anderson13}. The SED of LMXBs is assumed to follow power laws with slope $\Gamma=1.5$ {  ranging} from $0.3\,\rm keV$ to $8\,\rm keV$. The scaling relation between {  the} total mass of old stars and the X-ray luminosity of LMXBs is 
\begin{equation}
L_{\rm LMXB} \approx 10^{40}\frac{M_{\star}}{10^{11}{\rm M_{\odot}}}~{\rm erg~s^{-1}}.
\end{equation}

{  A single stellar particle in our simulation represents a stellar cluster rather than a single star. The spatial size of a stellar particle is set to $5\,\rm pc$, motivated by the typical size of giant molecular clouds (GMCs)/star {  clusters}. Beyond this radius, {  the radiation emitted from the stellar particle is modeled as a point source.} While within this radius, the radiation field is assumed to be uniform and equivalent to its value at $5\,\rm pc$.} Note, we neglect radiation {  originate} from BHs as the series of works {  focus} more on the impacts of local stellar radiation{  ;} the AGN radiation is not included in current works. Of course it should be explored in our future work. 

After obtaining the SED, we calculate the cooling and heating rate with radiation. For the primordial gas, the gas heating and cooling rate due to different processes is calculated by solving the ionization equilibrium equation \citep{katz96}. For the metal cooling and heating, the cooling and heating rate with different gas number density, temperature, redshift, the radiation flux of {  the star-forming} region and old stars, and metallicity is calculated with the photoionization code \textsc{Cloudy} \citep{ferland98} under the ionization equilibrium assumption, and is tabulated to feed the simulations. 

Once gas cooling/heating rate with radiation is obtained, we need to calculate the radiation intensity of the gas cells in {  the} simulation. For the UVB, the radiation field is assumed to be uniform. For the {  star-forming} region and old stars, we follow the treatment presented in \citet{kannan14} and \citet{woods15}. We identify stellar particles with a stellar age younger than 100 Myr as young stars and the rest as old stars. Each stellar particle is treated as a single radiation source, and the radiation from a single stellar particle is scaled with the mass of the stellar particle. After obtaining the SED and the luminosity of {  all the stellar particles}, the radiation intensity for any gas cells is calculated by summing up the radiation irradiated by {  all the stellar particles}. Here{  ,} we assume that, except for star-forming regions, the gas in a galaxy is optically thin in terms of radiation. To accelerate this calculation process, we use the octree to calculate the total radiation intensity for any gas cell. Since radiation flux from a source fade with $r^{-2}$ and eventually becomes weaker than the UVB, for each source{  ,} we only calculate its radiation within $r=1\,c{\rm Mpc}/h$.

\section{Results} \label{sec:results}

This section explores how the local stellar radiation influences the evolution, properties, and distribution of dwarf galaxies near the simulated proto MW-sized galaxies and {  compares} outcomes between the NoRad and StarRad simulations. In the following discussion, we denote {  a} ``luminous halo/dwarf'' as the halo/subhalo with at least one stellar particle.

\begin{figure*}
\centering
   \includegraphics[width=\textwidth]{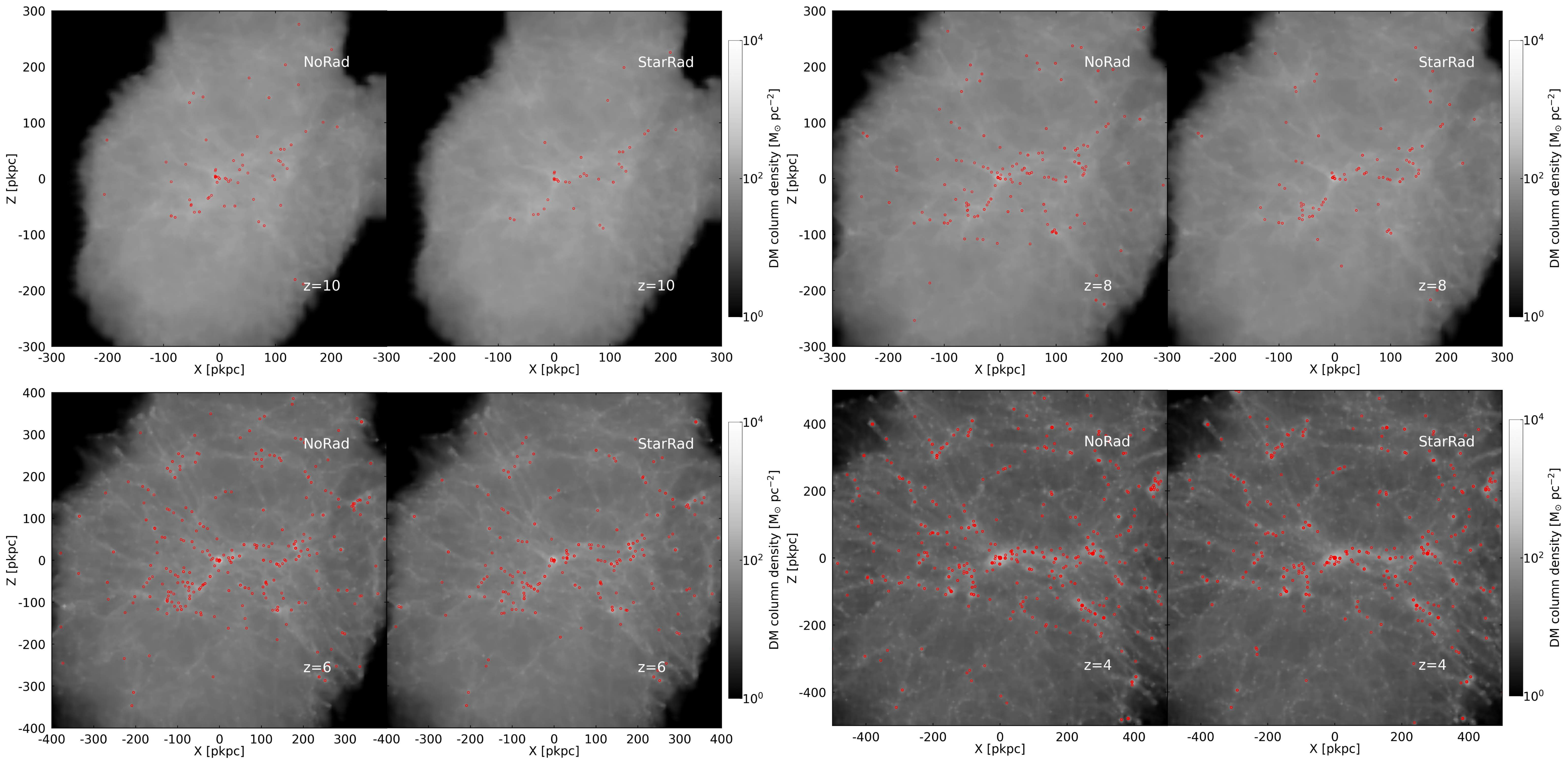}
   \caption{Dark matter density map of the Au16 simulations at z= 10, 8, 6, and 4, respectively. The red {  circles} indicate the luminous galaxies. In each panel, the left panel {  shows} the results of the NoRad run,  while the right panel {  show} the results in the StarRad run.}
    \label{SurfDM}
\end{figure*}

\begin{figure}
\centering
    \subfigure{\includegraphics[width=0.45\textwidth]{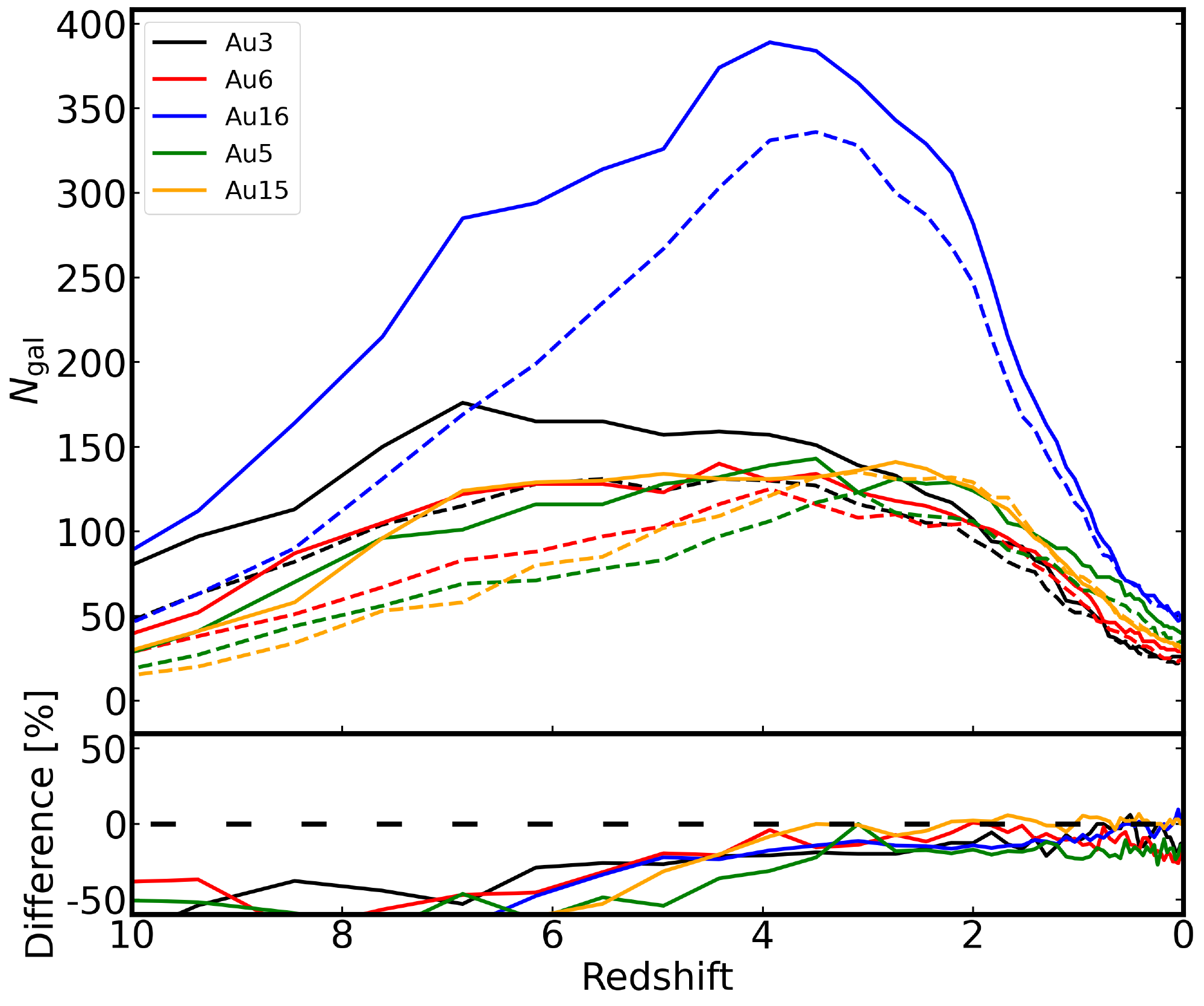}}
   
   \subfigure{\includegraphics[width=0.45\textwidth]{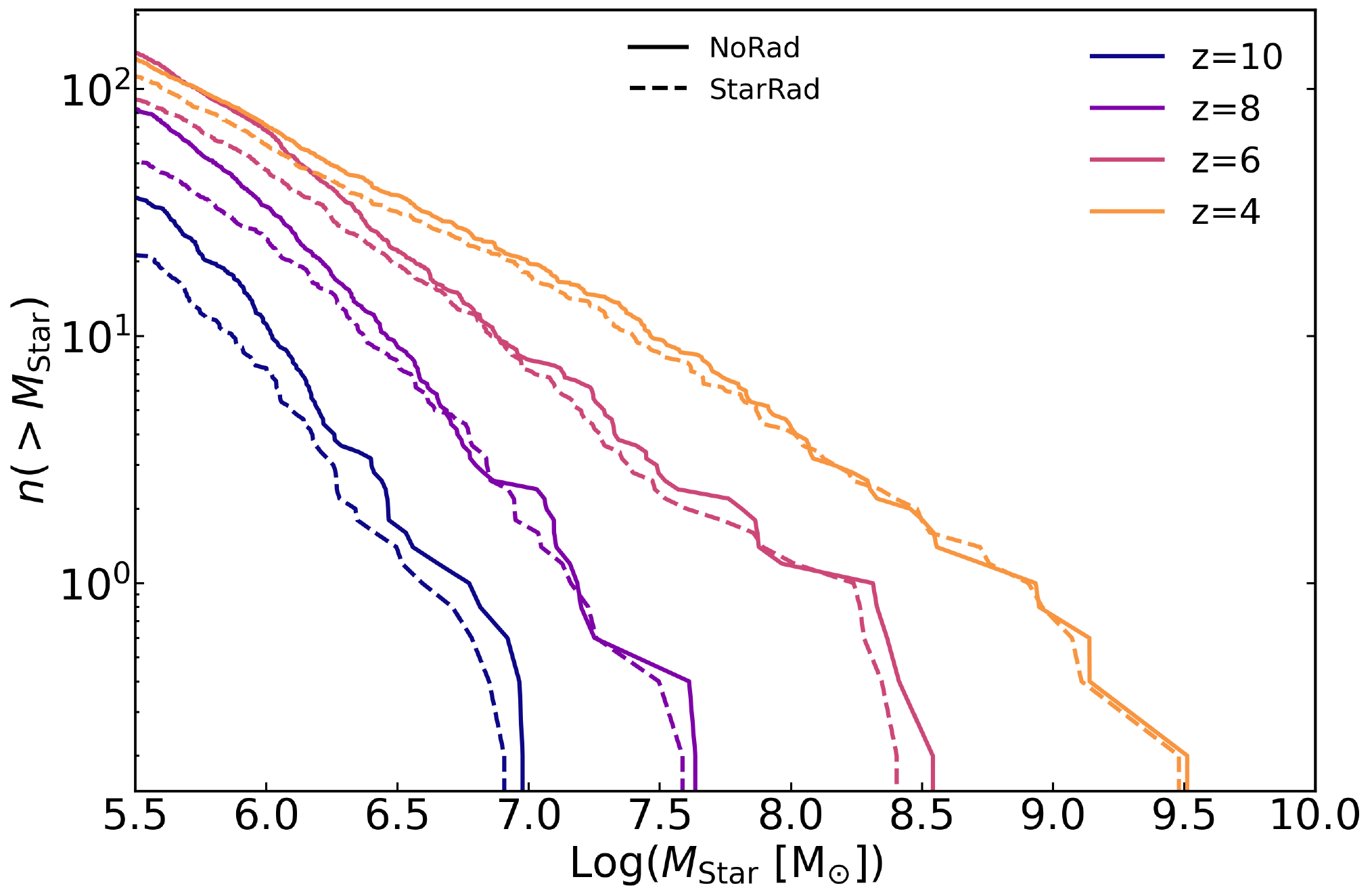}}
   \caption{{\it Upper panel:} The time evolution of the total number of luminous dwarfs within the selected radius around the {  dominant} galaxies. The solid lines show the results for the NoRad runs, while the dotted lines show the results for the StarRad runs. The bottom panel shows the relative difference between the NoRad and StarRad runs. {\it Lower panel:} The cumulative stellar mass function {  of} the NoRad and StarRad runs at different redshifts. The solid lines show the results for the NoRad runs, while the dotted lines show the results for the StarRad runs.}
    \label{timing}
\end{figure}

\begin{figure}
\centering
   \includegraphics[width=0.45\textwidth]{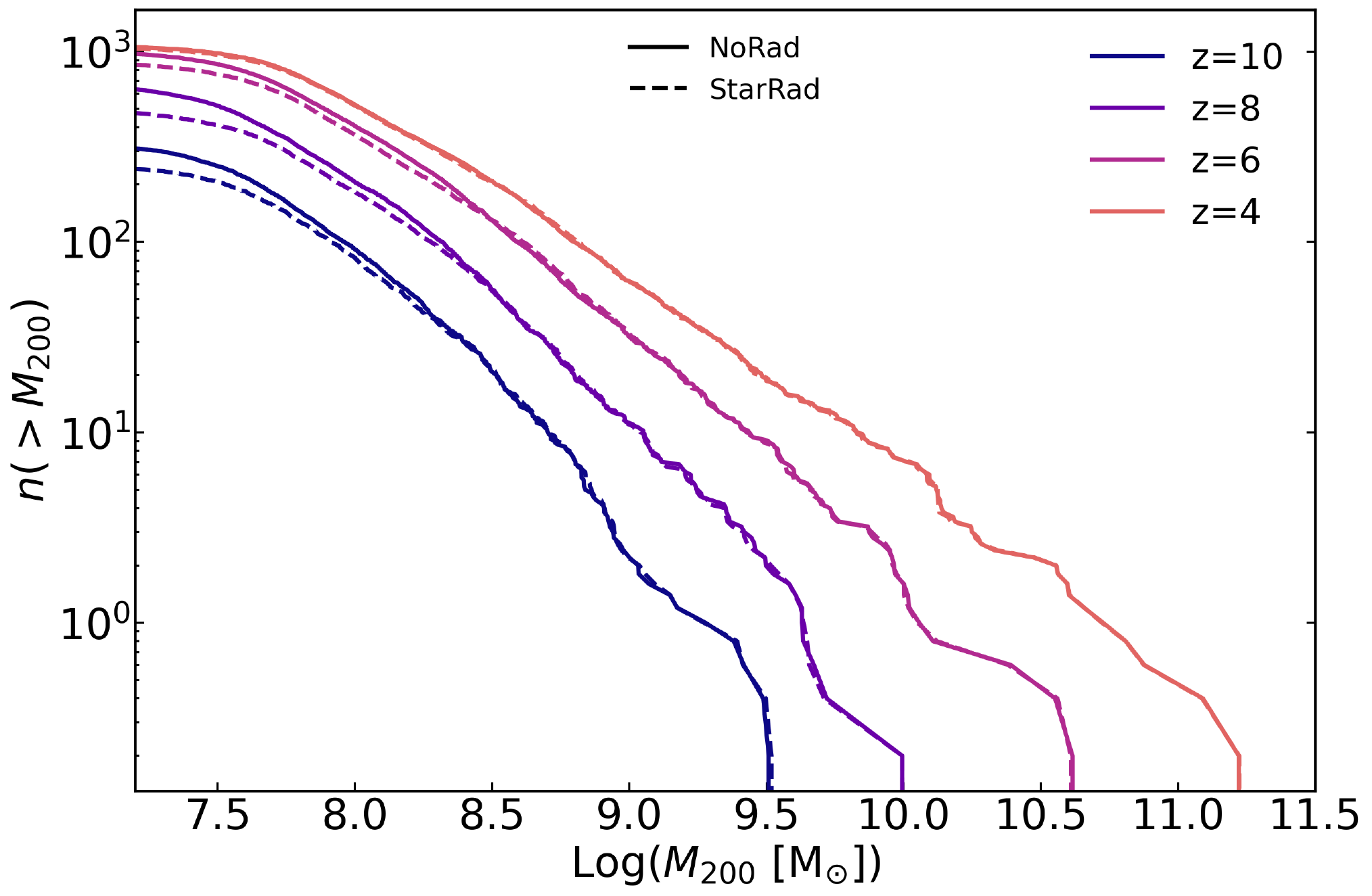}
   \caption{The cumulative halo mass function for the NoRad and StarRad simulations. The solid lines show the results for the NoRad runs, while the dotted lines show the results for the StarRad runs.}
    \label{mf}
\end{figure}

\subsection{The abundance of dwarf galaxies}\label{sec:spadis}

In Figure \ref{SurfDM}, we present {  a} visual impression of how the local stellar radiation {  affects} dwarf {  galaxy} formation in our simulations. For simplicity, we present results for the Au16 run, as it represents the most massive galaxy in our simulation suite and thus is expected to {  have the} most significant impact. In the plot, luminous dwarfs are overplotted as red dots {  on top of the} underlying DM density field of {  the high-resolution} region{ ;} results for the NoRad and StarRad cases {  at $4$ different redshifts are shown}. At first glance, comparing the results in the NoRad and StarRad runs, the number of luminous dwarfs in the StarRad run is lower than in the NoRad run from $z=10$ to $z=6$. However, the discrepancy decreases with decreasing redshift. At $z=4$, the difference between the NoRad and StarRad runs becomes negligible. These results indicate that the local stellar radiation can suppress {  the} formation of luminous halos at high redshift, and this effect becomes less pronounced with decreasing redshift. Note {  that} near the boundary of the {  high resolution} regions{ ,} our simulations may be contaminated by {  low-resolution} particles{ . Therefore} the results near the boundary are less reliable. However, because we focus on the relative difference, this {  does not affect} our conclusion. 

To quantify the above results, we display the time evolution of the total number of the luminous dwarfs in the {  high-resolution} regions of different simulated Auriga proto-galaxies in Figure \ref{timing}.  The sizes of the high-resolution regions in our simulations are approximately 200-300 $p$kpc at $z=10$ and 500-600 $p$kpc at $z=0$. These regions are smaller than the characteristic scale at which local stellar radiation becomes significantly attenuated by gas absorption (defined as the radius where the optical depth reaches 3). Therefore, the analysis of local stellar radiation effects within these regions can be considered reliable, as the radiation field remains largely unaffected by optical depth effects in this volume.

The relative difference of the total number between NoRad and StarRad galaxies is shown in the bottom panel of the same figure. Comparing the results in the NoRad and in StarRad runs, it is clear that local stellar radiation makes huge impact on the formation of dwarf galaxies at high redshift.
Before redshift $6$, the local stellar radiation reduces the number of dwarf galaxies by approximately 50\% on average across all our runs, with some variations depending on the specific halo configuration. After redshift $z=6$, the difference {  gradually decreases} with cosmic time and {  drops} to less than 10\% at $z\la3$. 

We also note that while the absolute number of dwarf galaxies varies between halos, the consistent trends observed in the relative difference between NoRad and StarRad simulations across the selected halos suggest that the impact of local stellar radiation is primarily determined by the existence of the local stellar radiation itself, rather than the formation history of the host halo. Therefore, our selection is not expected to introduce significant bias.

The bottom panel of Figure \ref{timing} {  shows} the time evolution of galaxy stellar mass function (GMF) from $z=10$ to $z=4$. At redshift $z=10$, the galaxy mass functions (GMFs) in the StarRad runs are suppressed by approximately 40\% across all mass scales relative to the NoRad runs. At lower redshifts, the discrepancies between the two become smaller and are only significant for galaxies with stellar masses $M_{\rm Star}<10^7\rm\,M_{\odot}$, where the suppression remains around 15-20\%.

Earlier works have demonstrated that the {  baryonic} processes also affect the structure formation of dark matters halos\citep{sawala2013, schaller2015, qin2017, zheng2024}.
It is interesting to see how the local stellar radiation {  affects} halo mass function (HMF). In Figure \ref{mf} we show the time evolution of HMF {  in high-resolution regions} in each simulation from $z=10$ to $z=4${ ;} the averaged results are shown. It is interesting that {  the} HMF in {  the} StarRad simulations is lower than in {  the} NoRad simulations for {  halo with masses} $M_{\rm 200}\la10^{8.5}\,{\rm M_{\odot}}$ at $z=10$ to $z\sim6$. 
This result can be explained by the indirect effect of photoionization heating from local stellar radiation. When local stellar radiation heats the gas to $T\sim10^{4.5}~\rm K$, the increased thermal pressure prevents gas from cooling and collapsing into low-mass halos. Since the formation and growth of dark matter halos are gravitationally coupled with baryons, the suppression of baryonic collapse also partially inhibits the growth and abundance of small dark matter halos before reionization.

At $z=4$, the difference between the NoRad and StarRad simulations {  becomes} negligible as the relative importance of the radiation decreases with redshift, which is consistent with the results shown in {  Figures} \ref{SurfDM} and \ref{timing}. 

\begin{figure*}
\centering
   \includegraphics[width=\textwidth]{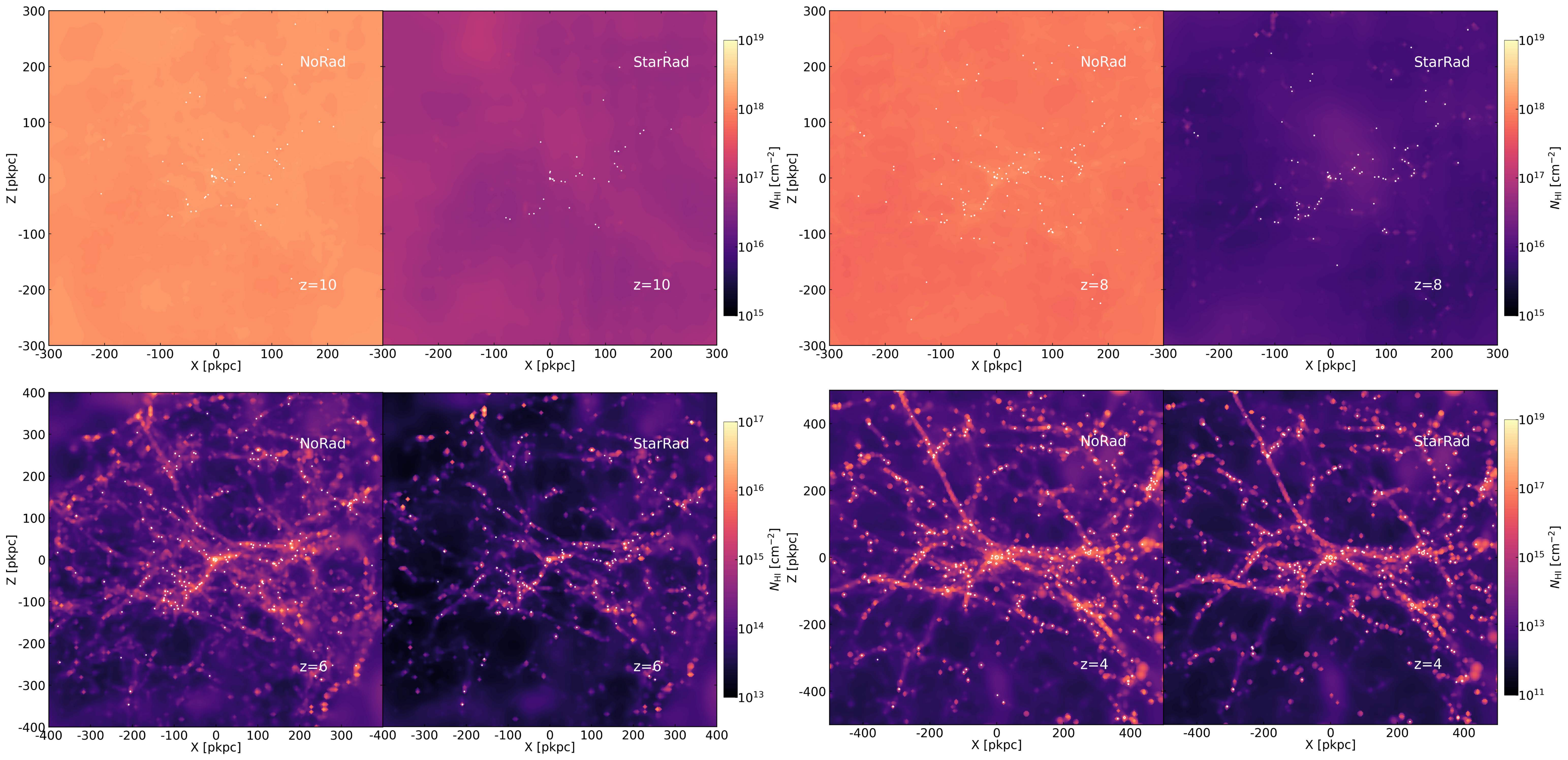}
   \caption{Similar to {  Figure} \ref{SurfDM}, but for HI column density at different redshifts in the Au13 NoRad and StarRad simulations. The white circles indicate luminous galaxies.}
    \label{coldensHI}
\end{figure*}

\begin{figure}
\centering
  \includegraphics[width=0.45\textwidth]{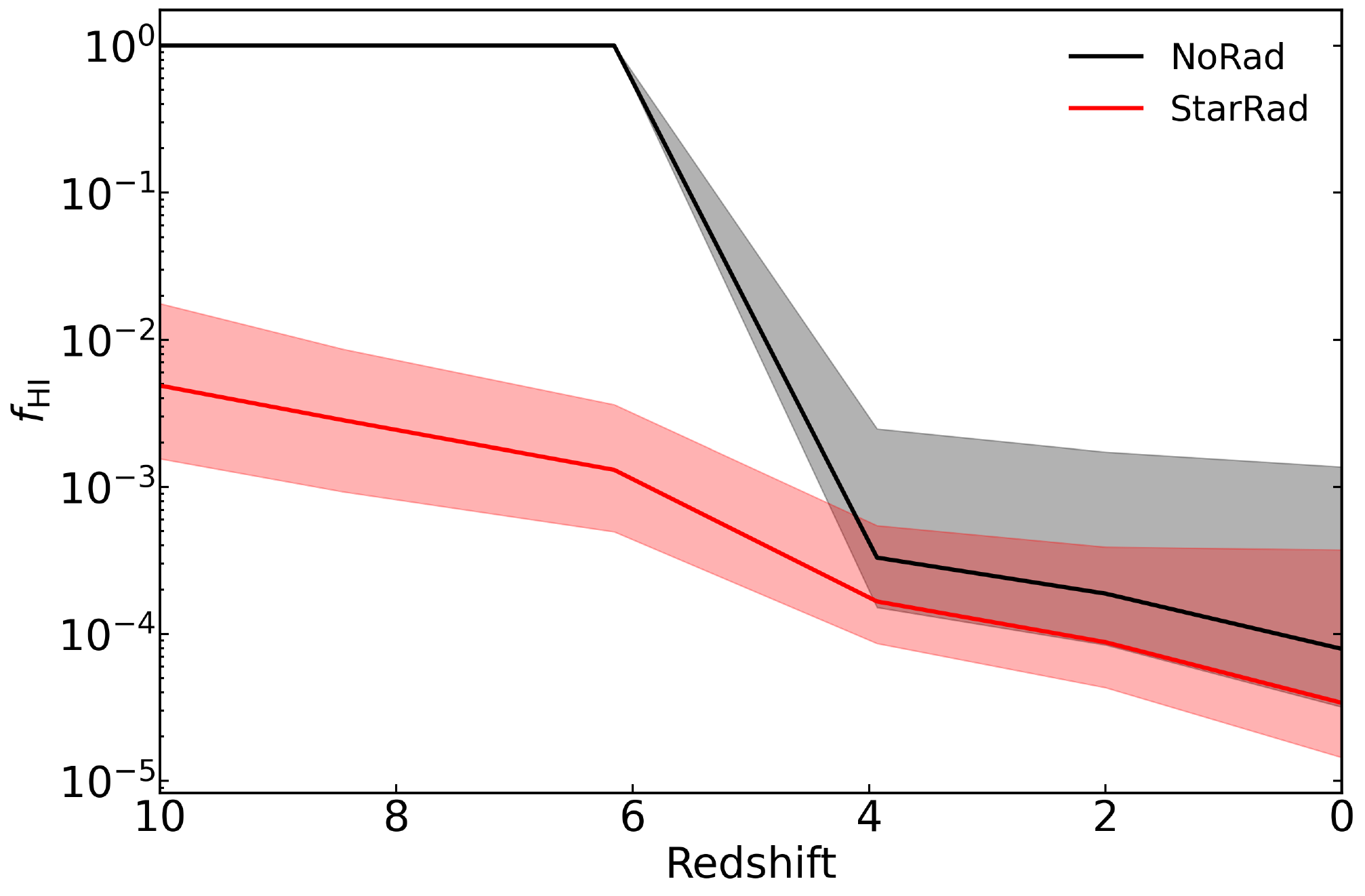}
   \caption{The median HI gas fraction within a range of the selected radius around each main halo as a function of redshift.  The shaded region {  represents 10th and 90th percentiles} of the values in each simulation set.}
   \label{mHIdis}
\end{figure}

\begin{figure}
\centering
   \includegraphics[width=0.45\textwidth]{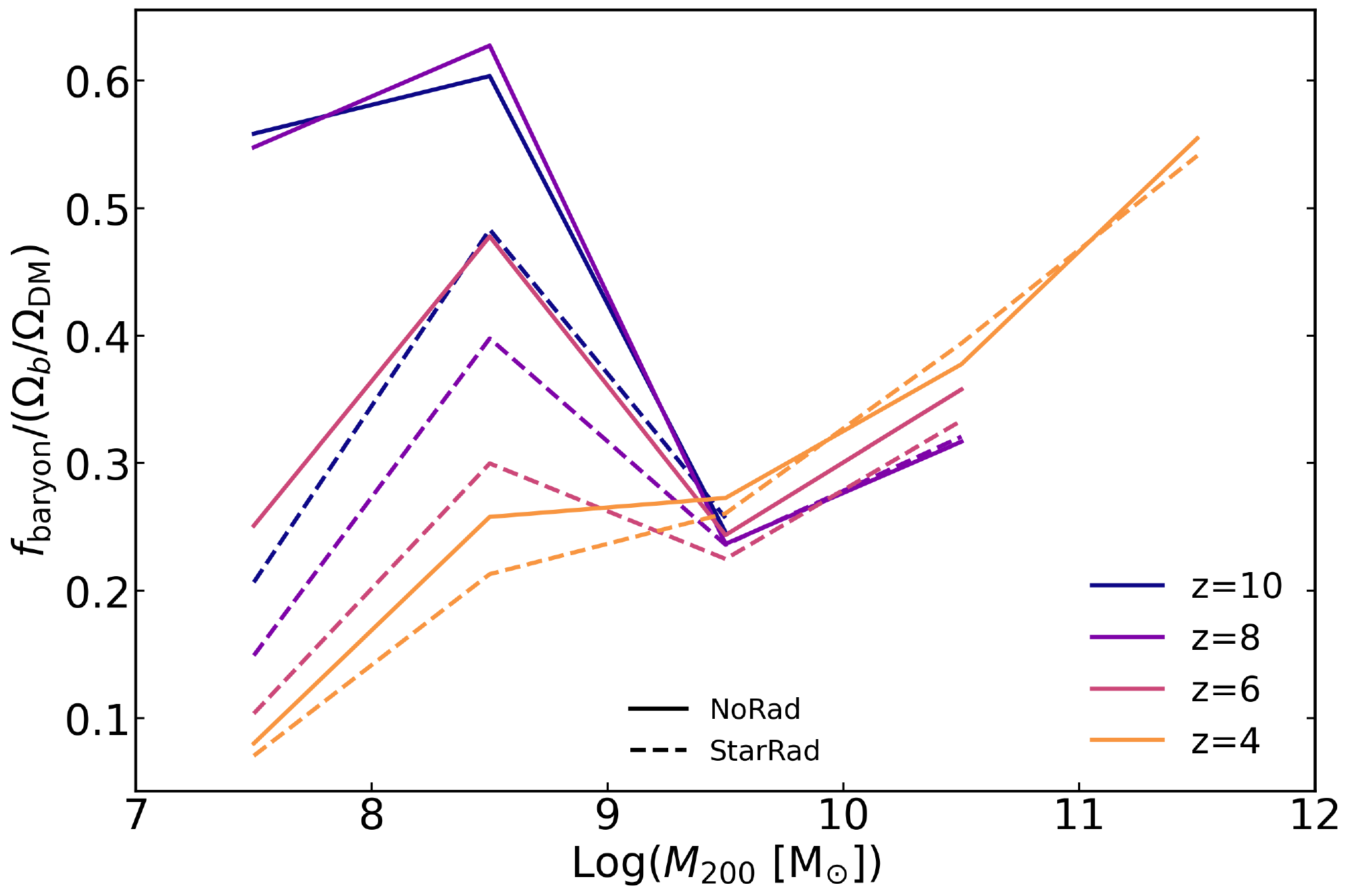}
   \caption{The median baryon fraction of dark matter halos as a function of halo mass at various redshifts.}
    \label{baryon}
\end{figure}

\subsection{Gas properties}\label{sec:gas}

In this subsection, we investigate the impact of the local stellar radiation on the gas properties. Figure \ref{coldensHI} presents HI column density maps of the Au13 simulations at z=10, 8, 6, and 4, respectively. In each panel, the left {  side} shows the results of the NoRad run, while the right one shows the {  results} of the StarRad run. The white {  circles} indicate {  the} virial radius of luminous dwarfs. Clearly{ ,} the HI column density of {  the} NoRad run is significantly higher than {  in the} StarRad run before {  the} redshift $z=6$, suggesting that the local stellar radiation can significantly reduce HI in the IGM. At redshift $4$, it is difficult to identify differences between the NoRad and StarRad runs.

The above results are quantified in Figure \ref{mHIdis}, which shows the HI gas fraction of our simulations as a function of redshift. The median values from five NoRad (black) and StarRad (red) simulations are shown, and the shaded regions {  represent 10th and 90th percentiles within each simulation set.} { In the NoRad simulations, by construction, the gas is completely neutral at $z>6$, whereas in the StarRad simulations, the presence of local stellar radiation results in partial ionization, leading to a lower HI fraction. This difference arises from the additional ionizing radiation in StarRad, which heats and ionizes the gas, reducing the HI column density. As the redshift decreases ($z<6$) and cosmic reionization {  is} complete, the HI fractions in both simulation sets converge.}

Another well-known effect of photoionization is that it can heat up gas to escape from low mass DM halos, leading to a significant reduction in the gas fraction, as shown by a few previous studies in {  the} context of the cosmic reionization \citep{barkana00, gnedin00, okamoto2008, wise08}. The same {  physical mechanism} should{ ,} of course{ ,} apply to the local stellar radiation. In Figure \ref{baryon}{ ,} we present the time evolution of the baryon fraction of dark matter halo as a function of halo mass in different runs{ ; results} for 4 different redshifts are shown. For dark matter halos with a halo mass $M_{\star}\la 10^{9.5}\,{\rm M_{\odot}}$, the baryon fraction is quite different between the ``NoRad'' and ``StarRad'' runs, the difference vanishes for the halos with higher masses.
In addition, the difference diminishes as redshift decreases, indicating the effects of the local stellar radiation for suppressing the gas accretion of the low-mass halo become weaker with decreasing redshift, which have similar trends as the results discussed above.

\begin{figure}
\centering
   \includegraphics[width=0.45\textwidth]{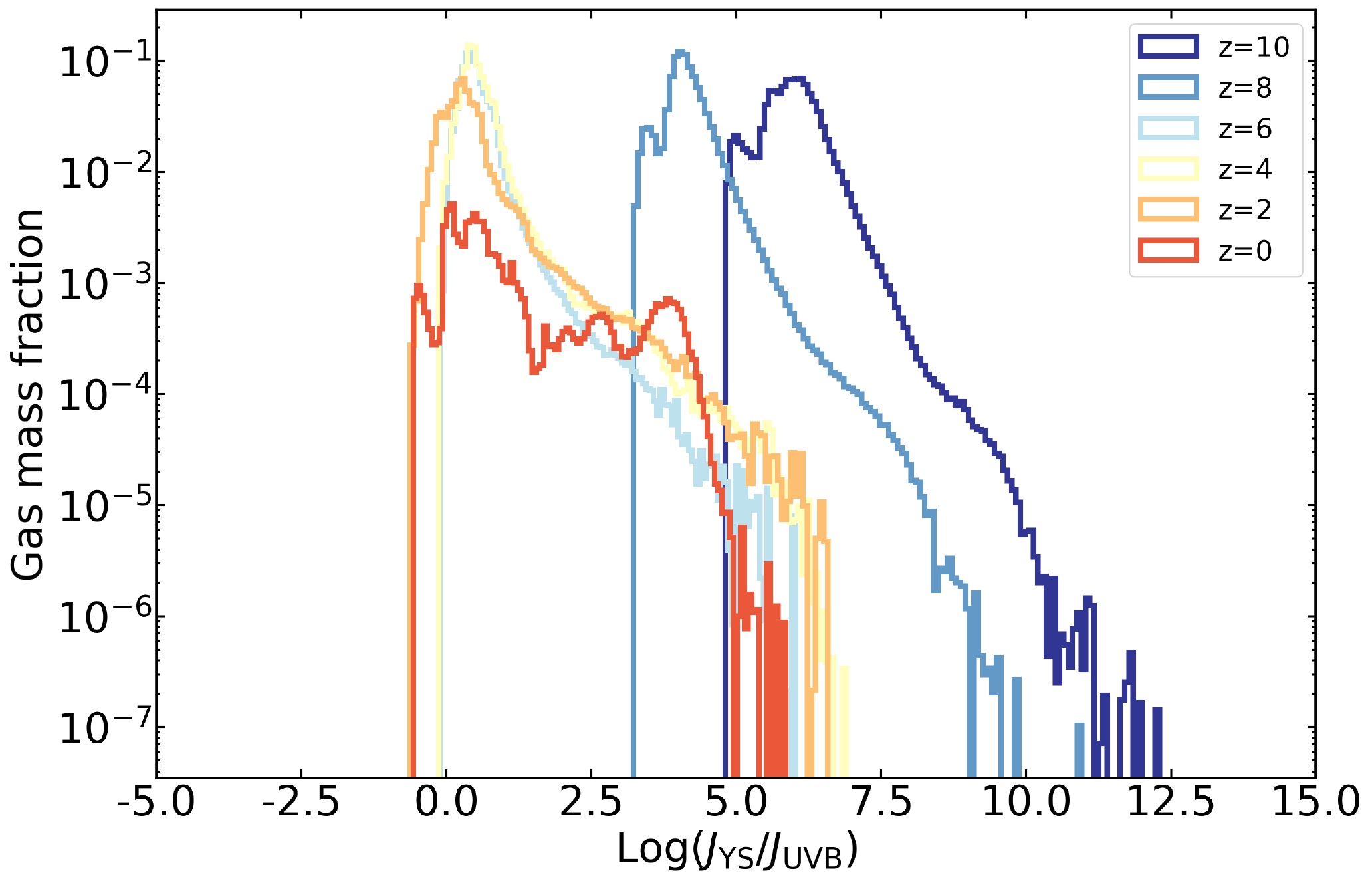}
   \caption{The gas mass distribution of the ratio between the intensity of the local stellar radiation and UVB within a range of the selected radius around the main halos {  where only high-resolution regions are included} at different redshifts. The gas is selected by the criterion that the temperature is lower than $10^{4.5}\,{\rm K}$.}
    \label{mgasdis}
\end{figure}

The impact of the uniform UVB on gas properties in {  low-mass} dark matter halos is well understood. To evaluate the impact of the local stellar radiation, it is interesting to compare its intensity to that of UVB. In Figure \ref{mgasdis}, we present the gas mass distribution of the ratio between radiation from {  star-forming} region and UVB within the high-resolution regions at various {  redshifts;} the results averaged from five StarRad simulations are shown. Based on the cooling/heating rate shown in \citet{zhu2024}, only the gas cell with temperature lower than $\sim10^{4.5}\,{\rm K}$ will be significantly affected by the radiation, thus only the gas cell with the temperature lower than $10^{4.5}\,{\rm K}$ is included in this figure. In the figure, we can see that before $z=8${ ,} the intensity of the local stellar radiation $J_{\rm YS}$ (since the local stellar radiation is dominated by young stellar region, we only take this part into account) is significantly higher than that of UVB ($J_{\rm UVB}$) in the simulation. After $z=6$, the peak of the ratio distribution is close to {  the} unit, indicating that after the epoch of reionization (at $z\sim 5.5$), the local stellar radiation and UVB {  become} comparable. The effects of the local stellar radiation become weak since its effects are only equivalent to approximately doubling UVB enhancement{ ,} and the fraction of the cold gas {  becomes} lower and lower after the epoch of reionization.

The sharp decline observed at the low-radiation end of the distribution (\(\log(J_{\rm YS}/J_{\rm UVB})\)) results from the limited physical extent of the high-resolution region used in our analysis. Since the local stellar radiation field approximately follows an inverse-square law (\(J \propto 1/r^2\)), the smallest radiation intensity within the analyzed volume is determined by the largest physical distance from the radiation sources, i.e., \(J_{\rm min} \sim L/r_{\rm max}^2\). {  The} high-resolution region is spatially limited, {  resulting} in a natural lower cutoff in the radiation field rather than a physical suppression of low-radiation regions.

The effects of the local stellar radiation become weak since its effects are only equivalent to approximately doubling the UVB intensity, and the fraction of cold gas decreases progressively after the epoch of reionization.

\subsection{The abundance of satellites at z=0} 

\begin{figure}
\centering
   \includegraphics[width=0.45\textwidth]{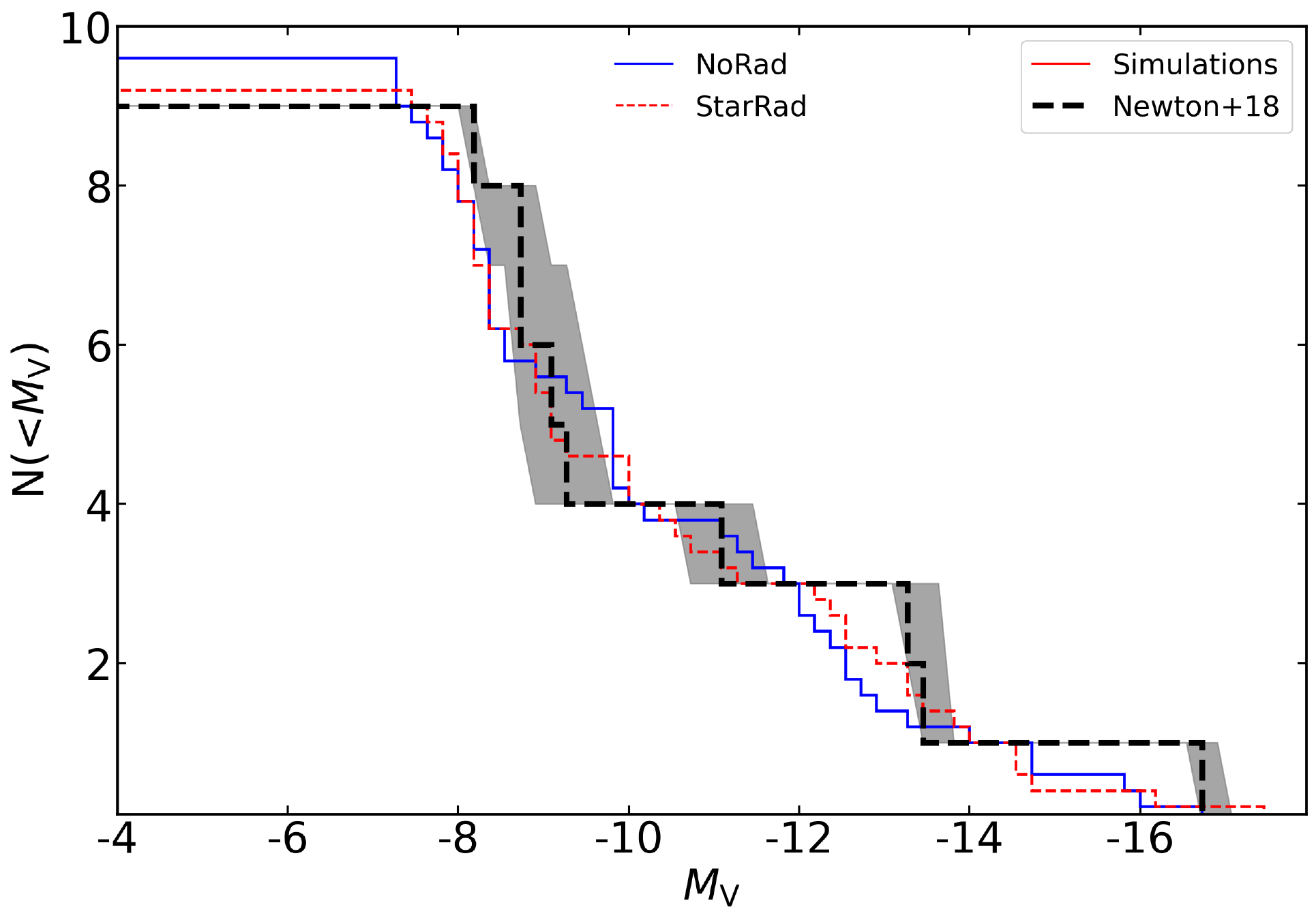}
   \caption{The V-band luminosity function of satellite galaxies in the main halos  at $z=0$. The thick dashed line represents the observational results from  \citet{newton18}. The {  gray-shaded} region represents the observational uncertainties derived from measurement errors, including photometric errors and distance uncertainties, following the data from \citet{mcconnachie12} and \citet{torrealba16}. The region corresponds to the 1$\sigma$ confidence interval.}
    \label{z0}
\end{figure}

In this {  subsection}, we explore how the local stellar radiation {  affects} the satellite population {  at} the present day. In Figure \ref{z0}, we compare the V-band luminosity function of the satellites in the StarRad and NoRad runs, {  where} the averaged values in each simulation {  set} are shown. For reference, we also {  include} observational {  results} from \citet{newton18}. Clearly, {  due to the numerical resolution limitations}, both simulation sets can only resolve classical satellites, {  which show nearly identical results in both cases}, suggesting the local stellar radiation has little effect on the brighter satellite galaxy population. This might not be surprising as the local stellar radiation only affects galaxy formation in low mass halos at high {  redshifts. The} brighter satellites are less affected by the local stellar radiation. 

\begin{figure}
\centering
   \includegraphics[width=0.45\textwidth]{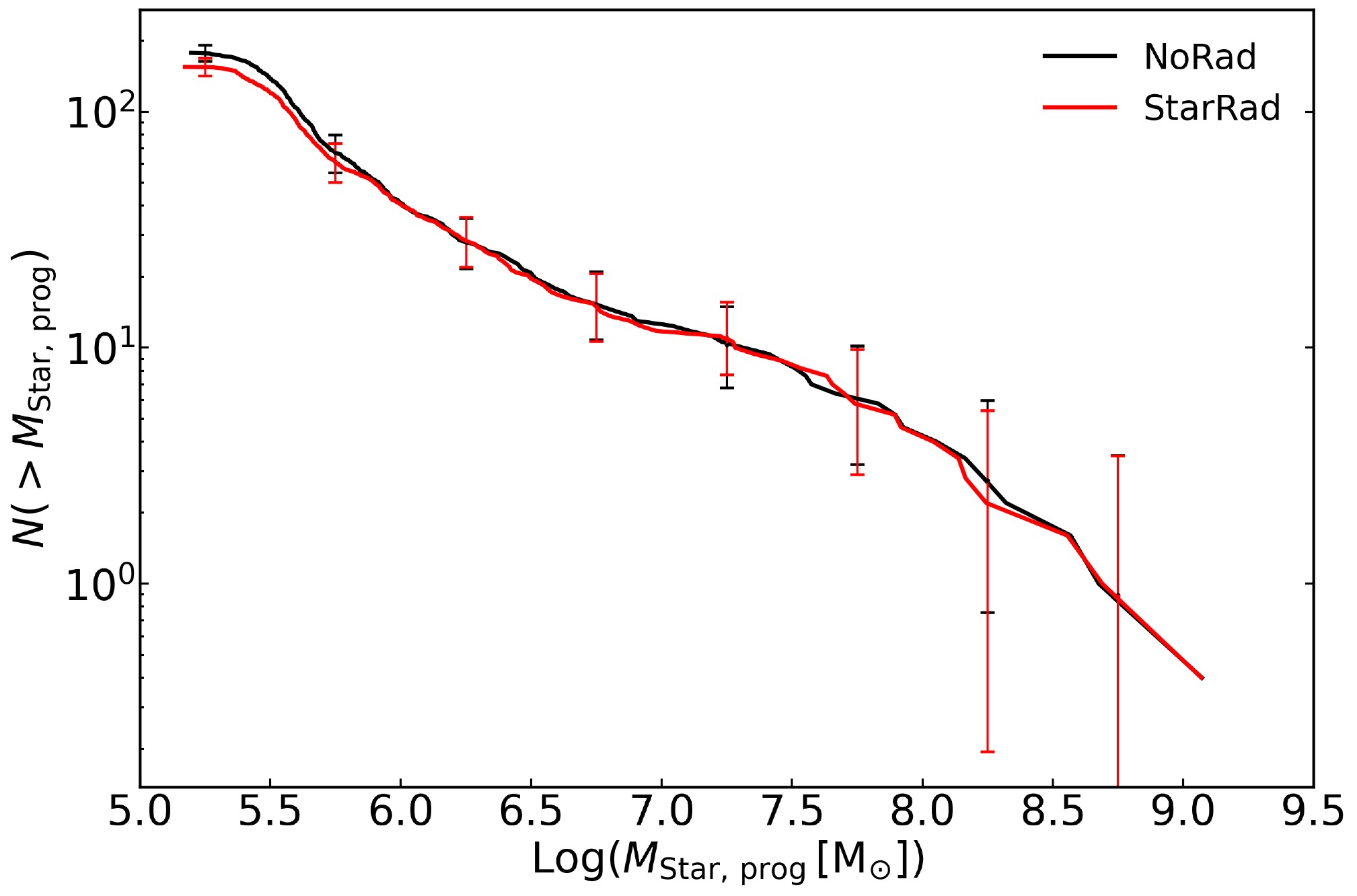}
   \caption{The accreted galaxy stellar mass function over the entire halo assembly. {  The error bars represent Poisson uncertainties calculated as $\sigma=\sqrt{N}$ , where N is the number of galaxies in each mass bin.}}
    \label{usmf}
\end{figure}

While {  local stellar radiation is expected to affect only} the faint end of {  the} satellite luminosity function, our simulations lack the necessary resolution to address this problem directly. However, we can {  still} {  gain some insight} by looking at the accreted dwarf galaxies at infall, which are possible progenitors of satellite galaxies at the present day. Once {  accreted into the central halos}, some of the progenitors are physically disrupted by tidal effects, while some of them may disappear only due to numerical resolution. Our simulations have sufficient resolution to resolve luminous dwarf galaxies { with $M_{\rm V}\la-7$} before infall, and hence{ ,} we can reliably assess the impact of local stellar radiation on the final satellite galaxy population using these progenitors. 
Figure \ref{usmf} {  shows} the progenitor mass function in the NoRad and StarRad runs. A progenitor is defined as a galaxy accreted into the virial radius of the {  central} halo, with its stellar mass recorded at the time of accretion. The averaged values are shown for both simulation sets. Poisson error bars, calculated as $\sigma=\sqrt{N}$ for each mass bin, are included to reflect statistical uncertainties. Both simulations show good agreement for the relatively massive progenitors ($M_{\rm Star, prog}\ga 10^8\,{\rm M_{\odot}}$), where the differences fall within the error bars. {  However, below this mass, the two runs diverge, with a 13\% difference observed at the lowest mass scale. This difference is within $\sim2\sigma$ of the Poisson uncertainty, indicating that it is not statistically significant. Therefore, while local stellar radiation may contribute to variations in the faint satellite population, this effect is within the expected statistical fluctuations.}

While the difference is not large, it should be taken into account when using the abundance of MW satellite galaxies to constrain dark matter models \citep{kennedy2014, bose2016, nadler2021, newton2021, hayashi2021, dekker2022}. 

\section{Summary and conclusion} \label{sec:summary}

In this work, we explore the impact of local stellar radiation on the formation and evolution of the dwarf galaxies {  surrounding MW-analogues} by using the simulations from \citetalias{zhu2024}. Our main findings can be summarized as follows.

The total {  number} of the dwarfs around the dominant galaxy is suppressed in the StarRad galaxies at $z\ga6$ {  by approximately 50\%}. However, after $z\sim6$, when reionization {  is} complete, the difference in the total amount of dwarf galaxies between the simulations with and without the local stellar radiation decreases with decreasing redshift, {  becoming} less than 10\% at $z\la3$. 

{  This result can be explained since the local stellar radiation as {  an} ionizing source can photoionize and heat up the IGM in the vicinity of proto-MW analogues before the reionization.} Consequently, the neutral gas fraction in the volume with local stellar radiation is significantly reduced before the reionization. {  Since the gas {  becomes} hotter, the low mass DM halos in {  the} proto-MW region are harder to capture gas to form stars. {  As} a result { ,} the abundance of {  the} dwarf {  galaxies} is significantly suppressed before reionization.} The median intensity of the local stellar radiation is larger than that of UVB by {  a factor of \(\sim 10^6\) at $z=10$ and \(\sim 3\) at $z=6$}, while they become comparable {  afterward}.

Although the impact of local stellar radiation on the formation of dwarf galaxies {  is significant} at high redshift,  this does not propagate to the present day. The V-band luminosity function of the resolved satellites in the halos of the simulations with and without the local stellar radiation is almost exactly the same. We further compare the accreted galaxy stellar mass function in both simulation sets and find {  that} the local stellar radiation can reduce the abundance of the faintest satellite by a factor of $13$ percent. {  However, this difference is within $\sim2\sigma$ of the Poisson uncertainty, indicating that it is not statistically significant.}

It is worth noting that there are still some limitations in the current treatment of local stellar radiation in our models. The radiative transfer is based on the optically-thin assumption, {  which neglects photon absorption and self-shielding effects in high-density regions. In a realistic medium, the optical depth $\tau$ increases with both gas density and distance from the radiation source, leading to significant attenuation of ionizing photons and reducing their effective reach. Since the optically-thin approximation ignores this attenuation, it tends to overestimate the impact of radiation by extending the ionized regions and causing stronger suppression of star formation, particularly in low-mass halos where the gravitational potential is shallow{ ,} and gas retention is more sensitive to radiative feedback.} Therefore, the results presented above should be considered as an upper limit on the effects of local stellar radiation. These considerations motivate us to implement a more accurate radiative transfer model, including optical depth effects, in future work.

\begin{acknowledgments}

We thank Shi Shao and Lizhi Xie for helpful discussions. The authors thank Volker Springel for providing the data. {  We thank the anonymous referee for their detailed and insightful feedback, which has significantly improved the clarity and scientific accuracy of this manuscript.} We acknowledge the supports from the National Natural Science Foundation of China (Grant No. 11988101) and the K. C. Wong Education Foundation. Data analysis were performed on the Freya compute cluster of MPA, operated by the Max Planck Computing and Data Facility.

\end{acknowledgments}

\bibliography{ref}{}
\bibliographystyle{aasjournal}



\end{document}